\documentclass[12pt]{article}

\usepackage{graphicx}
\begin{document}

\begin{center}
{\bf Magnetized black holes and nonlinear electrodynamics } \\
\vspace{5mm} S. I. Kruglov
\footnote{E-mail: serguei.krouglov@utoronto.ca}
\underline{}
\vspace{3mm}

\textit{Department of Chemical and Physical Sciences, University of Toronto,\\
3359 Mississauga Road North, Mississauga, Ontario L5L 1C6, Canada} \\
\vspace{5mm}
\end{center}
\begin{abstract}
A new model of nonlinear electrodynamics with two parameters is proposed. We study the phenomenon of vacuum birefringence, the causality and unitarity in this model. There is no singularity of the electric field in the center of point-like charges and the total electrostatic energy is finite. We obtain corrections to the Coulomb law at $r\rightarrow\infty$. The weak, dominant and strong energy conditions are investigated.
Magnetized charged black hole is considered and we evaluate the mass, metric function and their asymptotic at $r\rightarrow\infty$  and $r\rightarrow 0$. The magnetic mass of the black hole is calculated. The thermodynamic properties
and thermal stability of regular black holes are discussed. We calculate the Hawking temperature of black holes and show that there are first-order and second-order phase transitions. The parameters of the model when the black hole is stable are found.
\end{abstract}

\section{Introduction}

QED with quantum corrections modifies Maxwell's electrodynamics and becomes nonlinear electrodynamics (NLED) \cite{Heisenberg}. The phenomenon of vacuum birefringence takes place in this NLED in the presence of the external magnetic field. The effect of vacuum birefringence means that indexes of refraction are different for two orthogonal polarization states. This effect is now of great experimental interest \cite{Rizzo}, \cite{Valle}, \cite{Battesti}. Therefore, models of NLED that admit the phenomenon of vacuum birefringence are of definite interest.
In Born-Infeld (BI) electrodynamics \cite{Born} there is no the effect of birefringence,
but in the modified Born-Infeld electrodynamics with two parameters the birefringence phenomenon occurs \cite{Krug}.
For weak field limit NLED should be converted into Maxwell's electrodynamics.
Due to self-interaction of photons, for strong electromagnetic fields, classical electrodynamics may be modified \cite{Jackson}.
In BI electrodynamics and in other models of NLED \cite{Shabad} - \cite{Kruglov7} an upper bound on the electric field in the
center of charged particles exists and the total electromagnetic energy is finite. But in classical electrodynamics
there are problems of singularity of an electric field in the center and the infinite electromagnetic energy of charged particles. Such problems may be absent in NLED.
It is interesting that NLED coupled with general relativity (GR) can give the universe acceleration \cite{Garcia} - \cite{Kruglov4}. But electromagnetic fields in BI electrodynamics do not drive the universe to accelerate \cite{Novello1}. In addition, BI electrodynamics has the problem of causality \cite{Quiros}.
The black hole solutions in GR in the framework of different NLED were investigated in \cite{Oliveira} - \cite{Kruglov8} and in many other papers.
In this paper we propose and investigate a new model of NLED.

The structure of the paper is as follows. In section II we propose new model of NLED with two parameters $\beta$ and $\gamma$.
The phenomenon of vacuum birefringence is investigated. We show that at $\gamma =8\beta$ the birefringence effect disappears.
It is found the range of magnetic field when the causality and unitarity principles are satisfied
in our model. We show in section III that the dual symmetry is broken. It is proven that there is no singularity of the electric field strength at the origin for the point-like particles and the maximum electric field strength in the center is $E(0)=\sqrt{2}/\sqrt{\beta}$. The correction to the Coulomb law in the order of ${\cal O}(r^{-6})$ is obtained.
We demonstrate in section IV that the total electrostatic energy of point-like charges is finite. The scale invariance in our model is violated due to the presence of dimensional parameters. The weak, dominant and strong
energy conditions are satisfied.
We investigate the electric-magnetic duality and show that there is not a one to one correspondence between F and P fames.
NLED coupled with GR is studied in section V and we find the regular black hole solution. The mass, the metric function and their asymptotic at $r\rightarrow\infty$  and $r\rightarrow 0$ are evaluated. We also calculate the magnetic mass of black holes. There are no singularities of the Ricci scalar at $r\rightarrow\infty$  and $r\rightarrow 0$. In section VI we investigate the black holes thermodynamics and the thermal stability of charged black holes.
At different values of the parameter $c=2^{9/2}\sqrt{\beta}/(qG)$ there may be one, two or no horizons.
At definite conditions the first-order and second-order phase transitions in black holes take place.
In section VII we make a conclusion.

The use units in which the speed of light$=\hbar=1$, $\varepsilon_0=\mu_0=1$, and the metric signature of the Minkowski spacetime is $\eta=\mbox{diag}(-1,1,1,1)$.

\section{The model of nonlinear electrodynamics}

Let us introduce NLED with the Lagrangian density
\begin{equation}
{\cal L} = -\frac{{\cal F}}{(\beta{\cal F}+1)^2}+\frac{\gamma}{2}{\cal G}^2,
 \label{1}
\end{equation}
where the parameters $\beta$ and $\gamma$ have the dimensions of (length)$^4$ ($\beta{\cal F}$ and $\gamma{\cal G}$ are dimensionless), ${\cal F}=(1/4)F_{\mu\nu}F^{\mu\nu}=(\textbf{B}^2-\textbf{E}^2)/2$, ${\cal G}=(1/4)F_{\mu\nu}\tilde{F}^{\mu\nu}=\textbf{E}\cdot \textbf{B}$,
$F_{\mu\nu}=\partial_\mu A_\nu-\partial_\nu A_\mu$ is the strength of fields, and $\tilde{F}^{\mu\nu}=(1/2)\epsilon^{\mu\nu\alpha\beta}F_{\alpha\beta}$
is the dual tensor. The model described by Eq. (1) is the modification of the model
proposed in \cite{Kruglov}.
The model (1) possesses different behavior compared to [9] for strong electromagnetic fields
as in our model (at ${\cal G}=0$) $\lim_{{\cal F}\rightarrow \infty} {\cal L}=0$, but in model [9]
$\lim_{{\cal F}\rightarrow \infty} {\cal L}=-1/\beta$. In addition, in this paper we investigate the magnetized black holes and the black hole thermodynamics.
It will be shown that the model under consideration possesses attractive features.
Thus, we have the regular black hole solution. It should be noted that the first model possessing the regular black hole solution was proposed in \cite{Bardeen} and then investigated in \cite{Ayon}. But in that model the correspondence principle does not hold.
Other regular black hole models with nonlinear electromagnetic sources were considered in \cite{Ayon1}-\cite{Dymnikova}.
In the present model the first-order and second-order phase transitions in black holes take place.

The second term in Eq, (1), containing the scalar ${\cal G}$, has the same structure as a term in QED with one-loop
corrections \cite{Heisenberg}.  The correspondence principle in our model takes place because at $\beta{\cal F}\ll 1$ and
$\gamma{\cal G}\ll 1$ the Lagrangian density (1) approaches to Maxwell's
Lagrangian density ${\cal L}_{M}=- {\cal F}$. Thus, in the weak field limit the nonlinearity of field equations disappears.

\subsection{Vacuum birefringence}

QED was tested in the BMV experiment \cite{Rizzo} and in the PVLAS experiment \cite{Valle}
measuring the effect of vacuum birefringence. The phenomenon of vacuum birefringence occurs in QED due to one-loop corrections \cite{Heisenberg}. There is no vacuum birefringence in classical electrodynamics and in Born-Infeld (BI) \cite{Born} electrodynamics.
In generalized BI electrodynamics with two parameters \cite{Krug} the phenomenon of vacuum birefringence holds. Let us investigate the vacuum birefringence in the model of NLED (1). Assuming that $\beta {\cal F}\ll 1$ we obtain the Taylor series of the Lagrangian density (1) \begin{equation}
{\cal L} = -{\cal F}+4\beta {\cal F}^2-12\beta^2{\cal F}^3+{\cal O}\left((\beta {\cal F})^4\right)+\frac{\gamma}{2}{\cal G}^2.
 \label{2}
\end{equation}
One can compare Eq. (2) with the Lagrangian density investigated in \cite{Kruglov9} (see also \cite{Dittrich}),
\begin{equation}
{\cal L} =- \frac{1}{2}\left(\textbf{B}^2-\textbf{E}^2\right)+a\left(\textbf{B}^2-\textbf{E}^2\right)^2
+b\left(\textbf{E}\cdot\textbf{B}\right)^2.
\label{3}
\end{equation}
 We see, comparing Eqs. (2) and (3), that up to ${\cal O}\left((\beta{\cal F})^2\right)$, $a=\beta$,
$b=\gamma/2$. In accordance with the results of \cite{Kruglov9} the indexes of refraction $n_\perp$, $n_\|$ for two polarizations, perpendicular and parallel to the external magnetic induction field $\bar{B}$, are  as follows:
\begin{equation}
n_\perp=1+4a\bar{B}^2=1+4\beta\bar{B}^2,~~~~n_\|=1+b\bar{B}^2=1+\frac{\gamma}{2}\bar{B}^2.
\label{4}
\end{equation}
As a result, the phase velocities are $v_\perp=1/n_\perp<1$, $v_\|=1/n_\|<1$ and we have the phenomenon of vacuum birefringence
if $n_\perp\neq n_\|$.
The Cotton-Mouton (CM) effect \cite{Battesti} tells us that the difference in the indexes of refraction is given by
\begin{equation}
\triangle n_{CM}=n_\|-n_\perp=k_{CM}\bar{B}^2.
\label{5}
\end{equation}
From Eqs. (4) and (5) we obtain the CM coefficient $k_{CM}=\gamma/2-4\beta$.
The BMV and PVLAS experiments give the bounds
\[
k_{CM}=(5.1\pm 6.2)\times 10^{-21} \mbox {T}^{-2}~~~~~~~~~~(\mbox {BMV}),
\]
\begin{equation}
k_{CM}=(4\pm 20)\times 10^{-23} \mbox {T}^{-2}~~~~~~~~~~~~(\mbox {PVLAS}).
\label{6}
\end{equation}
From PVLAS experiment we find the bound on the parameters of our model $\gamma/2-4\beta\leq(4\pm 20)\times 10^{-23} \mbox {T}^{-2}$. If $\gamma=8\beta$ the effect of of vacuum birefringence is absent.
In QED, using quantum corrections, the bound on CM coefficient is $k_{CM}\leq 4.0\times 10^{-24}\mbox {T}^{-2}$ \cite{Rizzo}.
For strong magnetic fields the possible phenomenon of vacuum birefringence should be taken into consideration.

\subsection{The causality and unitarity principles}

For the healthy theory the general principles of causality and unitarity should be satisfied. Thus, the causality principle
guarantees that the group velocity of excitations over the background is less than the light speed. When the causality principle
holds tachyons will not appear. The unitarity principle requires that ghosts will be absent.
Both  requirements are formulated as \cite{Shabad2}
\[
 {\cal L}_{\cal F}\leq 0,~~~~{\cal L}_{{\cal F}{\cal F}}\geq 0,~~~~{\cal L}_{{\cal G}{\cal G}}\geq 0,
\] \begin{equation}
{\cal L}_{\cal F}+2{\cal F} {\cal L}_{{\cal F}{\cal F}}\leq 0,~~~~2{\cal F} {\cal L}_{{\cal G}{\cal G}}-{\cal L}_{\cal F}\geq 0,
\label{7}
\end{equation}
where ${\cal L}_{\cal F}\equiv\partial{\cal L}/\partial{\cal F}$, ${\cal L}_{\cal G}\equiv\partial{\cal L}/\partial{\cal G}$.
Making use of Eq. (1) we obtain
\[
{\cal L}_{\cal F}= \frac{\beta{\cal F}-1}{(1+\beta{\cal F})^3},~~~~ {\cal L}_{{\cal G}{\cal G}}=\gamma,
\]
\begin{equation}
{\cal L}_{\cal F}+2{\cal F} {\cal L}_{{\cal F}{\cal F}}=\frac{-3(\beta{\cal F})^2+8\beta{\cal F}-1}{(1+\beta{\cal F})^4},~~~~
{\cal L}_{{\cal F}{\cal F}}=\frac{2\beta(2-\beta{\cal F})}{(1+\beta{\cal F})^4}.
\label{8}
\end{equation}
From Eqs. (7) and (8) we find that the principles of causality and unitarity hold if the electromagnetic fields obey the inequality
\begin{equation}
0<\beta{\cal F}\leq \frac{4-\sqrt{13}}{3}\simeq 0.13.
\label{9}
\end{equation}
For the case $\textbf{E}=0$ this gives the restriction $ B\leq \sqrt{2(4-\sqrt{13})}/\sqrt{3\beta}\simeq
0.51/\sqrt{\beta}$.

\section{Field equations}

Euler-Lagrange equations lead to the equations of motion
\begin{equation}
\partial_\mu\left(\sqrt{-g}\left({\cal L}_{\cal F}F^{\mu\nu} +{\cal L}_{\cal G}\tilde{F}^{\mu\nu}\right) \right)=0.
\label{10}
\end{equation}
From Eqs. (1) and (10) we obtain field equations
\begin{equation}
 \partial_\mu\left(\sqrt{-g}\left(\frac{(\beta{\cal F}-1)F^{\mu\nu}}{(1+\beta{\cal F})^3}
+\gamma{\cal G}\tilde{F}^{\mu\nu}\right) \right)=0.
\label{11}
\end{equation}
One can find the electric displacement field $\textbf{D}=\partial{\cal L}/\partial \textbf{E}$,
\begin{equation}
\textbf{D}=\frac{1-\beta{\cal F}}{(1+\beta{\cal F})^3} \textbf{E}+\gamma {\cal G}\textbf{B}.
\label{12}
\end{equation}
We obtain the magnetic field from the relation $\textbf{H}=-\partial{\cal L}/\partial \textbf{B}$,
\begin{equation}
\textbf{H}= \frac{1-\beta{\cal F}}{(1+\beta{\cal F})^3}\textbf{B}-\gamma{\cal G}\textbf{E}.
\label{13}
\end{equation}
Eqs. (12) and (13) can be decomposed as \cite{Hehl}
\begin{equation}
D_i=\varepsilon_{ij}E^j+\nu_{ij}B^j,~~~~H_i=(\mu^{-1})_{ij}B^j-\nu_{ji}E^j.
\label{14}
\end{equation}
From Eqs. (12), (13) and (14) we find
\[
\varepsilon_{ij}=\delta_{ij}\varepsilon,~~~~(\mu^{-1})_{ij}=\delta_{ij}\mu^{-1},~~~~\nu_{ji}=\delta_{ij}\nu,
\]
\begin{equation}
\varepsilon=\frac{1-\beta{\cal F}}{(1+\beta{\cal F})^3},~~~~
\mu^{-1}=\varepsilon=\frac{1-\beta{\cal F}}{(1+\beta{\cal F})^3},~~~~\nu=\gamma {\cal G}.
\label{15}
\end{equation}
Field equations (11), using Eqs. (12) and (13), can be written in the form of nonlinear Maxwell's equations
\begin{equation}
\nabla\cdot \textbf{D}= 0,~~~~ \frac{\partial\textbf{D}}{\partial
t}-\nabla\times\textbf{H}=0.
\label{16}
\end{equation}
Equations (16) are nonlinear Maxwell's equations because $\varepsilon_{ij}$, $(\mu^{-1})_{ij}$, and $\nu_{ji}$
depend on electromagnetic fields.
From the Bianchi identity one obtains the second pair of nonlinear Maxwell's equations
\begin{equation}
\nabla\cdot \textbf{B}= 0,~~~~ \frac{\partial\textbf{B}}{\partial
t}+\nabla\times\textbf{E}=0.
\label{17}
\end{equation}
From Eqs. (12) and (13) we obtain the equality
\begin{equation}
\textbf{D}\cdot\textbf{H}=(\varepsilon^2-\nu^2)\textbf{E}\cdot\textbf{B}+2\varepsilon\nu{\cal F}.
\label{18}
\end{equation}
In our model the dual symmetry is broken because $\textbf{D}\cdot\textbf{H}\neq\textbf{E}\cdot\textbf{B}$ \cite{Gibbons}.
In BI electrodynamics and classical electrodynamics dual symmetry holds but in generalized BI electrodynamics \cite{Krug}
and QED with quantum corrections the dual symmetry is broken.

\subsection{The field of point-like charges}

The Maxwell's equation, for the point-like particle with the electric charge $Q$ ($\textbf{B}=0$), in Gaussian units is given by
\begin{equation}
\nabla\cdot \textbf{D}=4\pi Q\delta(\textbf{r})
\label{19}
\end{equation}
Taking into account Eq. (12), the solution to Eq. (19) is written as
\begin{equation}
\frac{E\left(1+\beta E^2/2\right)}{(1-\beta E^2/2)^3}=\frac{Q}{r^2}.
\label{20}
\end{equation}
If $r\rightarrow 0$ the solution to Eq. (20) is given by
\begin{equation}
E(0)=\sqrt{\frac{2}{\beta}}.
\label{21}
\end{equation}
Thus, there is no singularity of the electric field strength at the origin for the point-like charges.
The value (21) gives the maximum electric field at the center of charged particles. The same feature occurs
in BI electrodynamics, but in classical electrodynamics the electric field strength has the singularity at the origin of the point-like particles.
It is convenient to introduce unitless variables
\begin{equation}
x=\frac{\sqrt{2} r^2}{Q\sqrt{\beta}},~~~~y=\sqrt{\frac{\beta}{2}}E.
\label{22}
\end{equation}
Then Eq. (20) becomes
\begin{equation}
\frac{(1-y^2)^3}{y(1+y^2)}=x.
\label{23}
\end{equation}
The plot of the function $y(x)$ is represented by Fig. 1.
\begin{figure}[h]
\includegraphics[height=3.0in,width=3.0in]{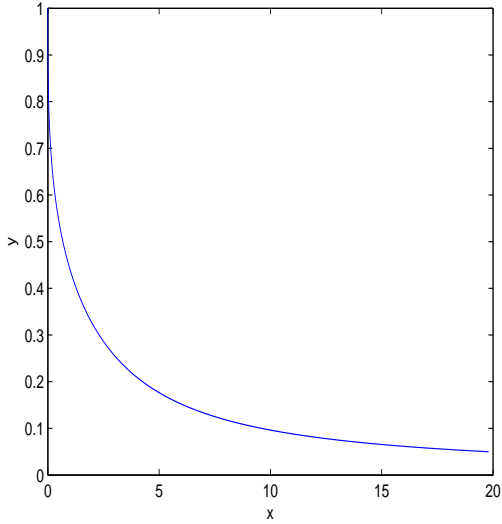}
\caption{\label{fig.1}The function  $y$ vs. $x$.}
\end{figure}
Numerical approximate real and positive solutions to Eq. (23) are given in Table 1.
\begin{table}[ht]
\caption{}
\centering
\begin{tabular}{c c c c c c c c c c  c}\\[1ex]
\hline \hline
$x$ & 1 & 2 & 3 & 4 & 5 & 6 & 7 & 8 & 9 & 10\\[0.5ex]
\hline
 $y$ & 0.440 & 0.324 & 0.256 & 0.209 & 0.176 & 0.152 & 0.133 & 0.118 & 0.106 & 0.096\\[0.5ex]
\hline
\end{tabular}
\end{table}
The Taylor series of the function $y(x)$ at $r\rightarrow \infty$ is
\begin{equation}
y=\frac{1}{x}-\frac{4}{x^3}+{\cal O}(x^{-5}).
\label{24}
\end{equation}
From Eqs. (22) and (24) we obtain the asymptotic value of the electric field  at $r\rightarrow\infty$
\begin{equation}
E(r)=\frac{Q}{r^2}-\frac{2\beta Q^3}{r^6}+{\cal O}(r^{-10}).
\label{25}
\end{equation}
The second term in the right side of Eq. (25) gives the correction to Coulomb's law.
Maxwell's electrodynamics is recovered at $\beta=0$ and we come to the Coulomb law $E=Q/( r^2)$ .
Thus, the electric field is finite at the center of the charged particles and singularities are absent.

\section{Energy-momentum tensor, dilatation current and energy of charges}

The symmetrical energy-momentum tensor can be obtained by varying the action on the metric tensor $g_{\mu\nu}$ \cite{Landau}.
This gives the expression
\begin{equation}
T_{\mu\nu}=\frac{2}{\sqrt{-g}}\frac{\partial (\sqrt{-g} {\cal L})}{\partial g^{\mu\nu}}.
\label{26}
\end{equation}
From Eqs. (1) and (26) we find the symmetrical energy-momentum tensor
\begin{equation}
T_{\mu\nu}=\frac{(\beta{\cal F}-1)F_\mu^{~\alpha}F_{\nu\alpha}}{(1+\beta{\cal F})^3}
+\gamma{\cal G}F_\mu^{~\alpha}\tilde{F}_{\nu\alpha}-g_{\mu\nu}{\cal L},
\label{27}
\end{equation}
From Eq. (27) one obtains the trace of the energy-momentum tensor
\begin{equation}
{\cal T}\equiv T_\mu^{~\mu}=\frac{8\beta{\cal F}^2}{(1+\beta{\cal F})^3}+2\gamma{\cal G}^2.
\label{28}
\end{equation}
At $\beta=\gamma=0$ we arrive at classical electrodynamics and the energy-momentum tensor becomes traceless.
We find the dilatation current and its divergence
\begin{equation}
D_{\mu}=x_\alpha T_{\mu\alpha},~~~~\partial_\mu D_{\mu}={\cal T}.
\label{29}
\end{equation}
As a result, the dilatation (scale) symmetry is violated because the dimensional parameters $\beta$, $\gamma$ are present.
The dilatation symmetry is  broken in NLED where there are dimensional parameters but in classical electrodynamics the dilatation symmetry occurs.

Let us calculate the total electrostatic energy of charged point-like particle. We obtain the energy density from Eq. (27)
\begin{equation}
\rho=T^{~0}_0=\frac{(1-\beta{\cal F})E^2}{(1+\beta{\cal F})^3} +\frac{{\cal F}}{(1+\beta{\cal F})^2}
+\frac{\gamma}{2}{\cal G}^2.
\label{30}
\end{equation}
In the case of electrostatics ($\textbf{B}=0$) the electric energy density (30) becomes
\begin{equation}
\rho_E=T^{~0}_0=\frac{2E^2(2+3\beta E^2)}{(2-\beta E^2)^3}.
\label{31}
\end{equation}
The total electrostatic energy of point-like particles ${\cal E}=\int_0^\infty \rho_E r^2dr$, making use of Eqs. (20), (22) and (31), is given by
\begin{equation}
{\cal E}=\frac{Q^{3/2}}{2^{7/4}\beta^{1/4}}\int_0^1 \frac{(1+3y^2)\sqrt{1-y^2}(3y^4+8y^2+1)dy}{\sqrt{y}(1+y^2)^{5/2}}
\simeq \frac{1.144Q^{3/2}}{\beta^{1/4}}.
\label{32}
\end{equation}
Thus, the total electrostatic energy of point-like particles is finite.
One can speculate that the electron mass is the total electrostatic energy \cite{Born1}, \cite{Rohrlich}, \cite{Spohn}. The point of view that the electron can be considered classically as a charged object was proposed by Dirac \cite{Dirac1}.

\subsection{Energy conditions}

Let us study the energy conditions that are of importance for viability of the theory.
The weak energy condition (WEC) \cite{Hawking} guarantees that the energy density is positive for any local
observer, and it is given by
\begin{equation}
\rho\geq 0,~~~\rho+p^m\geq 0 ~~~~ (m=1,~2,~3),
\label{33}
\end{equation}
where $\rho$ is the energy density and $p^m$ are principal pressures $p^m=-T_m^{~m}$ ($m=1,2,3$) and there is no summation in the index $m$. We will consider two cases: I) $\textbf{B}=0$, $\textbf{E}\neq0$ and II) $\textbf{E}=0$, $\textbf{B}\neq0$.

I) $\textbf{B}=0$, $\textbf{E}\neq0$.

It follows from Eq. (31) that $\rho_E\geq 0$. It should be mentioned that $E<E_{max}=\sqrt{2/\beta}$. From Eq. (27) we obtain
\begin{equation}
p^m_E=-T_m^{~m}=\frac{E^2}{2(1-\beta E^2/2)^2}-\frac{E^mE_m(1+\beta E^2/2)}{(1-\beta E^2/2)^3}.
\label{34}
\end{equation}
Then from Eqs. (31) and (34) one finds
\begin{equation}
\rho_E+p^m_E=\frac{(E^2-E_mE^m)(1+\beta E^2/2)}{(1-\beta E^2/2)^3}\geq 0.
\label{35}
\end{equation}
Thus, WEC is satisfied for the electric field values $E<E_{max}$.
The dominant energy condition (DEC) \cite{Hawking}, which shows that the speed of sound is less than the speed of light, is defined as
\begin{equation}
\rho\geq 0,~~~\rho+p^m\geq 0,~~~\rho-p^m\geq 0~~(m=1,~2,~3).
\label{36}
\end{equation}
It follows from Eqs. (33) and (36) that DEC includes WEC.
From Eqs. (31) and (34) at $\textbf{B}=0$, $\textbf{E}\neq 0$ one obtains
\begin{equation}
\rho_E-p^m_E=\frac{ E^mE_m(1+\beta E^2/2)+\beta E^4}{(1-\beta E^2/2)^3}\geq 0.
\label{37}
\end{equation}
As a result, DEC holds.
The strong energy condition (SEC) \cite{Hawking}, that defines the acceleration, is given by
\begin{equation}
\rho+\sum_{m=1}^3p^m\geq 0.
\label{38}
\end{equation}
With the help of Eqs. (31) and (34), we find
\begin{equation}
\rho_E+\sum_{m=1}^3p^m_E=\frac{E^2}{(1-\beta E^2/2)^2}\geq 0,
\label{39}
\end{equation}
and, therefore, SEC is satisfied. One can find the pressure, for the case $\textbf{B}=0$, from the relation
\begin{equation}
p_E= {\cal L}+\frac{E^2}{3}{\cal L}_{\cal F} =\frac{1}{3}\sum_{m=1}^3p_E^m.
\label{40}
\end{equation}
Thus, SEC (39) can be formulated as $\rho_E+3p_E\geq 0$. Then, for the case of NLED coupled with GR, SEC due to Friedmann's
equation tell us that electrically charged universe decelerates.
Let us investigate the second case.

II) $\textbf{E}=0$, $\textbf{B}\neq 0$.

 From Eqs. (30) and (27) we obtain
\begin{equation}
\rho_M=\frac{B^2}{2(1+\beta B^2/2)^2}\geq 0,
\label{41}
\end{equation}
\begin{equation}
p^m_M=\frac{(B^2-B^mB_m)(1-\beta B^2/2)}{(1+\beta B^2/2)^3}-\frac{B^2}{2(1+\beta B^2/2)^2}.
\label{42}
\end{equation}
Making use of Eqs. (41) and (42) one finds
\begin{equation}
\rho_M+p^m_M=\frac{(B^2-B^mB_m)(1-\beta B^2/2)}{(1+\beta B^2/2)^3}.
\label{43}
\end{equation}
Therefore, WEC is satisfied if $ B\leq \sqrt{2/\beta}$.
We obtain from Eqs. (41) and (42)
\begin{equation}
\rho_M-p^m_M=\frac{2B^2-B^mB_m+B^mB_m(\beta B^2/2)}{(1+\beta B^2/2)^3},
\label{44}
\end{equation}
and, as a result, DEC holds. One finds
\begin{equation}
\rho_M+\sum_{m=1}^3p^m_M=\frac{B^2-(3/2)\beta B^4)}{(1+\beta B^2/2)^3}.
\label{45}
\end{equation}
Therefore, SEC is satisfied at $B\leq \sqrt{2}/\sqrt{3\beta}\simeq 0.82/\sqrt{\beta}$.
For our case $\textbf{E}=0$ the pressure is
\begin{equation}
p_M={\cal L}-\frac{2B^2}{3}{\cal L}_{\cal F}=\frac{1}{3}\sum_{m=1}^3p_M^m.
\label{46}
\end{equation}
So, Friedmann's equation shows that magnetized universe decelerates at $B\leq \sqrt{2}/\sqrt{3\beta}\simeq 0.82/\sqrt{\beta}$, but
it accelerates at $B\geq \sqrt{2}/\sqrt{3\beta}\simeq 0.82/\sqrt{\beta}$.

\subsection{Electric-magnetic duality}

By virtue of a Legendre transformation \cite{Garcia1} one can consider an alternative form of NLED. We imply that $\textbf{B}=0$ (${\cal G}=0$), $\textbf{E}\neq0$. Let us introduce the tensor $P_{\mu\nu}$ and its invariant $P$,
\[
P_{\mu\nu}=\frac{1}{2}{\cal L}_{\cal F}F_{\mu\nu}=\frac{F_{\mu\nu}(\beta{\cal F}-1)}{2({\beta\cal F}+1)^3},
\]
\begin{equation}
P=P_{\mu\nu}P^{\mu\nu}=\frac{{\cal F}(\beta{\cal F}-1)^2}{({\beta\cal F}+1)^6},
\label{47}
\end{equation}
where ${\cal F}=-E^2/2$.
The Hamilton-like variable is given by
\begin{equation}
{\cal H}=2{\cal F}{\cal L}_{\cal F}-{\cal L}=\frac{{\cal F}(3\beta{\cal F}-1)}{({\beta\cal F}+1)^3}.
\label{48}
\end{equation}
It is easy to verify that the ${\cal H}$ is equal to the energy density (31), ${\cal H}=\rho_E$. One can check that the relations hold as follows:
 \begin{equation}
{\cal L}_{\cal F}{\cal H}_P=1,~~~~P{\cal H}_P^2={\cal F},~~~~{\cal L}=2P{\cal H}_P-{\cal H},
\label{49}
\end{equation}
where
\begin{equation}
{\cal H}_P=\frac{\partial {\cal H}}{\partial P}
=\frac{({\cal F}+1)^3(8\beta{\cal F}-3(\beta{\cal F})^2-1)}{11({\beta\cal F})^2-3(\beta{\cal F})^3-9\beta{\cal F}+1}.
\label{50}
\end{equation}
The plot of the function $\beta P(\beta{\cal F})$ is represented in Fig. 2.
\begin{figure}[h]
\includegraphics[height=3.0in,width=3.0in]{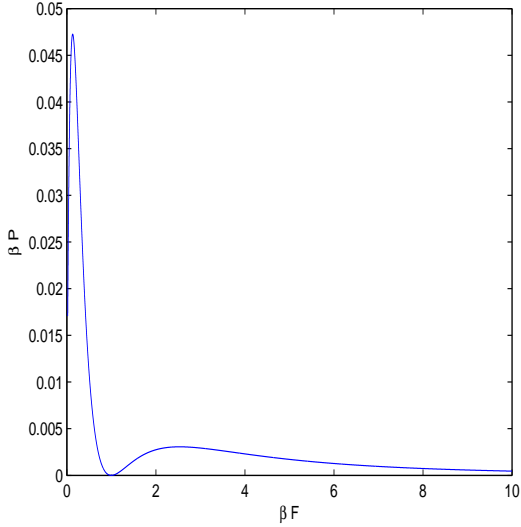}
\caption{\label{fig.2}The plot of the function $\beta P(\beta{\cal F})$.}
\end{figure}
Fig. 2 shows that the function ${\cal F}(P)$ is not a monotonic function. Therefore, there is not a one to one correspondence between ${\cal F}$ and $P$  frames  \cite{Bronnikov}. As a result, the electric-magnetic duality between two frames is broken. An electric solution in ${\cal F}$ frame with the Lagrangian density ${\cal L}({\cal F})$ does not possess a counterpart in the $P$ frame by the substitution ${\cal F}\rightarrow P$, ${\cal L}\rightarrow {\cal H}$, $F_{01}\rightarrow P_{23}$, and conversely. In Maxwell's electrodynamics we have ${\cal L}={\cal H}=-{\cal F}=-P$. In weak field regime, $\beta{\cal F}\ll 1$, both models, (1) and (48) are converted into the Maxwell theory ${\cal L}=-{\cal F}$.

\section{Magnetized black hole}

The action of our model of NLED in GR is
\begin{equation}
I=\int d^4x\sqrt{-g}\left(\frac{1}{2\kappa^2}R+ {\cal L}\right).
\label{51}
\end{equation}
Here $\kappa^2=8\pi G\equiv M_{Pl}^{-2}$, $G$ is Newton's constant, $R$ is the Ricci scalar, and $M_{Pl}$ is the reduced Planck mass.
The Einstein equation obtained from Eq. (51) is given by
\begin{equation}
R_{\mu\nu}-\frac{1}{2}g_{\mu\nu}R=-\kappa^2T_{\mu\nu}.
\label{52}
\end{equation}
Varying action (51) by electromagnetic potentials we find the equation of motion for electromagnet fields
\begin{equation}
\partial_\mu\left(\sqrt{-g}(F^{\mu\nu}{\cal L}_{\cal F}+\tilde{F}^{\mu\nu}{\cal L}_{\cal G})\right)=0.
\label{53}
\end{equation}
Let us investigate the static magnetic black hole solutions to Eqs. (52) and (53).
Bronnikov shown \cite{Bronnikov} that for pure magnetic field, when spherical symmetry holds, the invariant is
 ${\cal F}=q^2/(2r^4)$, where $q$ is a magnetic charge. In the case of the spherical symmetry, the line element is given by
\begin{equation}
ds^2=-f(r)dt^2+\frac{1}{f(r)}dr^2+r^2(d\vartheta^2+\sin^2\vartheta d\phi^2),
\label{54}
\end{equation}
with the metric function
\begin{equation}
f(r)=1-\frac{2GM(r)}{r}.
\label{55}
\end{equation}
The mass function is defined as
\begin{equation}
M(r)=\int_0^r\rho_M(r)r^2dr=m-\int^\infty_r\rho_M(r)r^2dr,
\label{56}
\end{equation}
were $m=\int_0^\infty\rho(r)r^2dr$ is the magnetic mass of the black hole. The magnetic energy density, found from Eq. (30),
is
\begin{equation}
\rho_M=\frac{2q^2r^4}{(2r^4+\beta q^2)^2}.
\label{57}
\end{equation}
From Eqs. (56) and (57) one obtains the mass function
\[
M(r)=\frac{q^{3/2}}{2^{3/4}\beta^{1/4}}\int_0^x \frac{x^6dx}{(x^4+1)^2}
=\frac{q^{3/2}}{2^{23/4}\beta^{1/4}}\biggl[3\sqrt{2}\ln\frac{x^2-\sqrt{2}x+1}{x^2+\sqrt{2}x+1}
\]
\begin{equation}-\frac{8x^3}{x^4+1}
+6\sqrt{2}\left(\arctan(1+\sqrt{2}x)-\arctan(1-\sqrt{2}x)\right) \biggr],
\label{58}
\end{equation}
where $x=2^{1/4}r/(\sqrt{q}\beta^{1/4})$.
From Eq. (58) we find the black hole magnetic mass
\begin{equation}
 m=M(\infty)=\frac{3\pi q^{3/2}}{2^{17/4}\beta^{1/4}}\simeq \frac{0.495q^{3/2}}{\beta^{1/4}}.
\label{59}
\end{equation}
Taking into consideration Eqs. (55) and (58) one obtains the metric function
\[
f(x)=1-\frac{1}{cx}\biggl[3\sqrt{2}\ln\frac{x^2-\sqrt{2}x+1}{x^2+\sqrt{2}x+1}
\]
\begin{equation}-\frac{8x^3}{x^4+1}
+6\sqrt{2}\left(\arctan(1+\sqrt{2}x)-\arctan(1-\sqrt{2}x)\right) \biggr],
\label{60}
\end{equation}
where $c=2^{9/2}\sqrt{\beta}/(Gq)$.
From Eq. (60) we find the asymptotic of the metric function at $x\rightarrow 0$ ($r\rightarrow 0$)
\begin{equation}
f(x)=1-\frac{1}{c}\left(\frac{32}{7}x^6-\frac{64}{11}x^{10}+\frac{32}{5}x^{14}+{\cal O}(x^{18})\right),
\label{61}
\end{equation}
or in the equivalent form
\begin{equation}
f(r)=1-\frac{4Gr^6}{7\beta^2q^2}+\frac{16Gr^{10}}{11\beta^3q^4}-\frac{16Gr^{14}}{5\beta^4q^6}+{\cal O}(r^{18}).
\label{62}
\end{equation}
It follows from Eq. (62) that we have the regular solution for magnetized black hole within our NLED because $\lim_{r\rightarrow 0} f(r)=1$.
One can find the asymptotic of the metric function at $x\rightarrow \infty$ ($r\rightarrow \infty$) from Eq. (60)
\begin{equation}
f(x)=1-\frac{1}{c}\left(\frac{6\sqrt{2}\pi}{x}-\frac{32}{x^2}+\frac{64}{5x^6}+{\cal O}(x^{-10})\right).
\label{63}
\end{equation}
With the use of Eq. (59) and the definition $c=2^{9/2}\sqrt{\beta}/(Gq)$ we obtain from (63)
\begin{equation}
f(r)=1-\frac{2Gm}{r}+\frac{G q^{2}}{r^2}-\frac{\beta Gq^4}{5r^6}+{\cal O}(r^{-10}).
\label{64}
\end{equation}
Equation (64) shows that we have the RN solution with corrections in the order of ${\cal O}(r^{-6})$.
If $r\rightarrow \infty$ one has $f(\infty)=1$ and the spacetime becomes flat. At $\beta=0$  NLED becomes
Maxwell's electrodynamics and solution (64) is the RN solution.

We obtain event $x_+$ and internal Cauchy $x_-$ horizons by solving the equation $f(r)=0$. Table 2 represents  horizons for different parameters $c$.
\begin{table}[ht]
\caption{Event $x_+$ and internal Cauchy $x_-$ horizons}
\centering
\begin{tabular}{c c c c c c c c c c c}\\[1ex]
\hline \hline 
$c$ & 1 & 1.5 & 2 & 2.5 & 3 & 3.5 & 4 & 4.5 & 5 & 5.5\\[0.5ex]
\hline 
 $x_+$ & 25.397 & 16.477 & 11.995 & 9.284 & 7.455 & 6.124 & 5.095 & 4.254 & 3.516 & 2.733\\[0.5ex]
\hline
 $x_-$ & 0.850 & 0.942 & 1.024 & 1.104 & 1.186 & 1.274 & 1.374 & 1.495 & 1.661 & 1.977\\[1ex]
\hline
\end{tabular}
\end{table}
The plot of the function $c(x)$ (at $f(x)=0$) is represented in Fig. 3.
\begin{figure}[h]
\includegraphics[height=3.0in,width=3.0in]{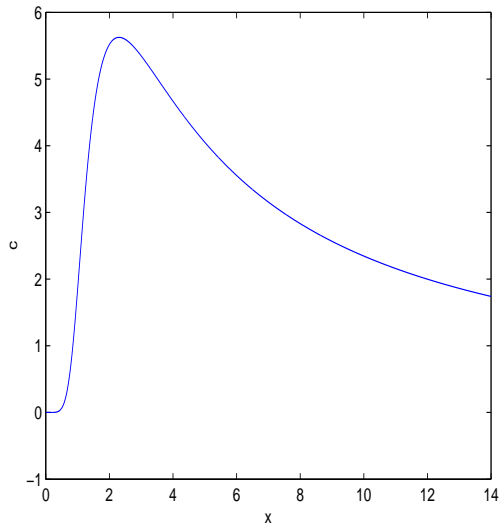}
\caption{\label{fig.3} The plot of the function $c(x)$.}
\end{figure}
Fig. 3 shows that it can be the regular black hole solution at $0<c<5.625$ with two horizons or the extremal black hole solution at $c\simeq 5.625$ with one horizon, or there can be no horizons at $c>5.625$ corresponding to the particle-like solution (similar to the magnetic monopole). The plot of the function $f(x)$ for $c=9, 5.625, 4$ is represented in Fig. 4.
\begin{figure}[h]
\includegraphics[height=3.0in,width=3.0in]{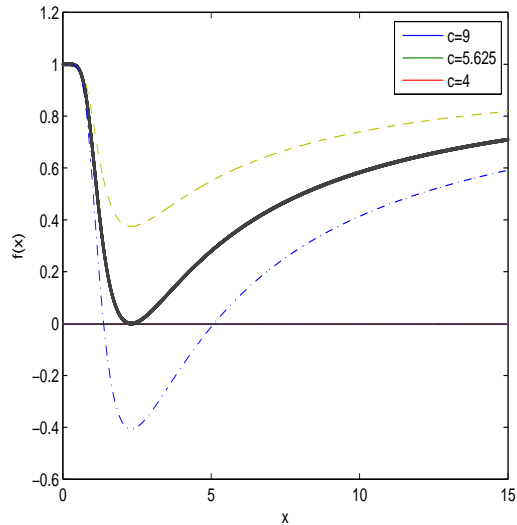}
\caption{\label{fig.4}The plot of the function $f(x)$. The dash curve corresponds to $c=9$, the solid (thick) curve is for
$c=5.625$, and the dashed-dot curve corresponds to $c=4$.}
\end{figure}

We obtain the Ricci scalar from Eqs. (28) and (52)
\begin{equation}
R=\kappa^2{\cal T}=\frac{16\kappa^2\beta q^4r^{3/2}}{(2r^4+\beta q^2)^3}.
\label{65}
\end{equation}
At $r\rightarrow \infty$ and at $r\rightarrow 0$ the Ricci scalar goes to zero, $R\rightarrow 0$, and therefore,
there are no singularities of the Ricci scalar.

\section{The black hole thermodynamics}

Let us study the black holes thermodynamics and the thermal stability of charged black holes. For this purpose we will calculate the temperature of the black hole. The Hawking temperature is given by
\begin{equation}
T_H=\frac{\kappa_S}{2\pi}=\frac{f'(r_+)}{4\pi},
\label{66}
\end{equation}
where $\kappa_S$ being the surface gravity and $r_+$ is the event horizon. From Eqs. (55) and (56) we obtain the relations as follows:
\begin{equation}
f'(r)=\frac{2 GM(r)}{r^2}-\frac{2GM'(r)}{r},~~~M'(r)=r^2\rho,~~~M(r_+)=\frac{r_+}{2G}.
\label{67}
\end{equation}
From Eqs. (57), (58), (66), and (67) one finds the Hawking temperature
\begin{equation}
T_H=\frac{1}{2^{7/4}\pi\sqrt{q}\beta^{1/4}}\left(\frac{1}{x_+}-\frac{32x_+^5}{c(x_+^4+1)^2}\right),
\label{68}
\end{equation}
where
\[
c=\frac{1}{x_+}\biggl[3\sqrt{2}\ln\frac{x_+^2-\sqrt{2}x_++1}{x_+^2+\sqrt{2}x_++1}
\]
\begin{equation}
-\frac{8x_+^3}{x_+^4+1}
+6\sqrt{2}\left(\arctan(1+\sqrt{2}x_+)-\arctan(1-\sqrt{2}x_+)\right) \biggr],
\label{69}
\end{equation}
and $x_+=\left(2/\beta q^2\right)^{1/4}r_+ , c=2^{9/2}\sqrt{\beta}/(Gq)$.
The plot of the function $T_H\sqrt{q}\beta^{1/4}$ is given in Fig. 5.
\begin{figure}[h]
\includegraphics[height=3.0in,width=3.0in]{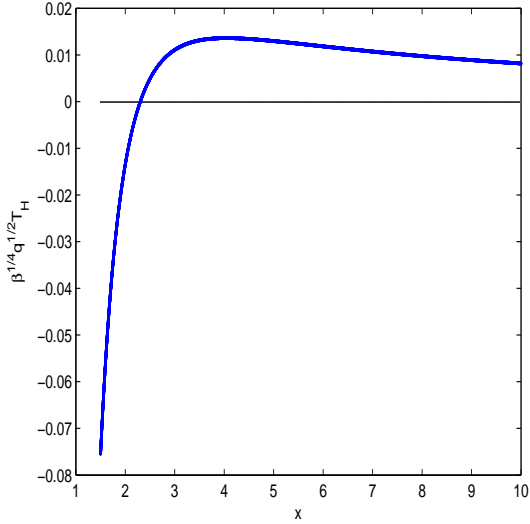}
\caption{\label{fig.5}The plot of the function $T_H\sqrt{q}\beta^{1/4}$ vs $x_+$.}
\end{figure}
The first-order phase transition takes place if the temperature and heat capacity change the sign.
The black hole is in the unstable state when the temperature is negative.
If the heat capacity is singular in some point, it corresponds to the second-order phase transition.
At $x_+\simeq 2.303$ ($r_+\simeq 1.94\sqrt{q}\beta^{1/4}$) the temperature becomes zero, $T_H=0$ and, it corresponds to the first-order black hole phase transition.

The black hole with zero temperature corresponds to the extremal one and differs significantly from other cases. The extreme and non-extreme black holes are different via certain semiclassical effects. But quantum effects change the spacetime geometry close to the event horizon of a black hole and, therefore, the surface gravity and temperature
are altered \cite{Bardeen1}. As a result, in physically realistic cases the macroscopic zero temperature black hole solutions do not exist. Zero temperature static black hole solutions in the framework of semiclassical theory of gravity can be considered as non-physical and they cannot join smoothly to the Reissner-Nordstr\"{o}m solution. It is impossible to construct a macroscopic zero temperature black hole that is near to zero temperature. The third law of black hole thermodynamics tells that a system cannot be reduced to zero temperature in a finite number of operations, i.e. a nonextremal black hole cannot become extremal \cite{Israel}. Thus, the extreme black hole with zero temperature cannot be produced by any physical processes. In this paper we explore only classical approach and do not consider quantum effects.

If $x_+< 2.303$ the Hawking temperature is negative and the black hole is unstable.
Making use of Eq. (69) we find the constant $c\simeq 5.625$ corresponding to the first-order phase transition ($x_+\simeq 2.303$). Then one obtains the critical values of the parameters corresponding to this horizon
\begin{equation}
\beta=\frac{(cqG)^2}{2^9}\simeq 0.062 q^2G^2,~~~m=\frac{0.495 q^{3/2}}{\beta^{1/4}}\simeq \frac{0.993q}{\sqrt{G}},~~~
T_H\simeq 0.
\label{70}
\end{equation}
Let us explore the entropy satisfying the Hawking area low
$S=A/(4G)=\pi r_+^2/G$. The heat capacity at the constant charge becomes
\begin{equation}
C_q=T_H\left(\frac{\partial S}{\partial T_H}\right)_q=\frac{T_H\partial S/\partial r_+}{\partial T_H/\partial r_+}=\frac{2\pi r_+T_H}{G\partial T_H/\partial r_+}.
\label{71}
\end{equation}
The temperature possesses the maximum at $x_+\simeq 4$ ($r_+\simeq 3.92\sqrt{q}\beta^{1/4}$). Then $\partial T_H/\partial r_+=0$ and the heat capacity (71) diverges indicating on the phase transition of the second-order \cite{Israel}. Therefore, we have the second-order phase transition at $x_+\simeq 4$.
The plots of the function $C_qG/(q \sqrt(\beta))$ vs $x_+$ are represented in Figs. 6 and 7.
\begin{figure}[h]
\includegraphics[height=3.0in,width=3.0in]{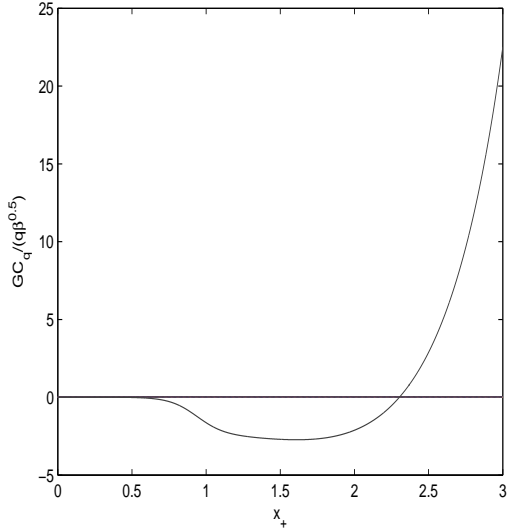}
\caption{\label{fig.6}The plot of the function $C_qG/(q \sqrt\beta)$ vs $x_+$.}
\end{figure}
\begin{figure}[h]
\includegraphics[height=3.0in,width=3.0in]{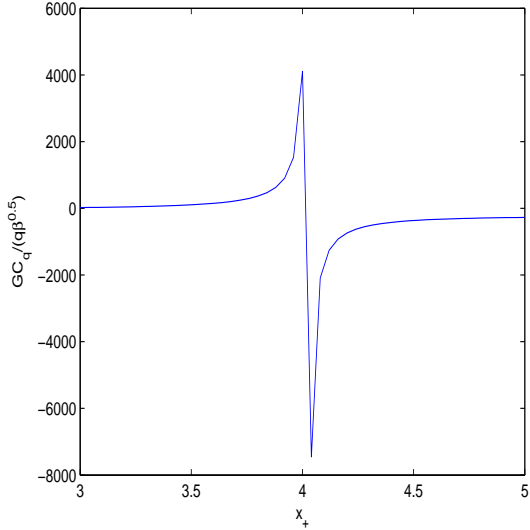}
\caption{\label{fig.7}The plot of the function $C_qG/(q \sqrt\beta)$ vs $x_+$.}
\end{figure}
By virtue of Eq. (69) we obtain the constant $c\simeq 4.667$ corresponding to the second-order phase transition. One can obtain the critical values of the
parameters which correspond to the horizon $x_+\simeq 4$
\begin{equation}
\beta=\frac{(cqG)^2}{2^9}\simeq 0.043 q^2G^2,~~~m=\frac{0.495 q^{3/2}}{\beta^{1/4}}\simeq \frac{1.087q}{\sqrt{G}},~~~
T_H\simeq \frac{0.0136}{q\sqrt{G}}.
\label{72}
\end{equation}
As a result, the parameters, which correspond to
the phase transitions, can be expressed via the magnetic charge of the black hole $q$ and Newton's constant $G$.
When the horizon $r_+$ is greater than the critical value $r_+ \simeq 3.92\sqrt{q}\beta^{1/4}$ the black hole becomes unstable.
Thus, the black hole within our model is stable in the range $1.94\sqrt{q}\beta^{1/4}<r_+<3.92 \sqrt{q}\beta^{1/4}$.

 \section{Conclusion}

We have proposed a new model of NLED with two parameters $\beta$ and $\gamma$ which for weak fields is converted to Maxwell's electrodynamics. The birefringence
phenomenon holds if $\gamma\neq 8\beta$ but otherwise the effect of birefringence disappears similar to classical electrodynamics.
It is known that in QED the birefringence phenomenon takes place due to quantum corrections. We have shown that, for the case of the presence of pure magnetic field, at $ B\leq \sqrt{2(4-\sqrt{13})}/\sqrt{3\beta}\simeq
0.51/\sqrt{\beta}$ the principles of causality and unitarity occur. The
dual symmetry is broken in our model as well as in QED with loop corrections. It was demonstrated that there is no singularity of the electric field in the center of
point-like particles and the maximum electric field strength is $E(0)=\sqrt{2}/\sqrt{\beta}$. The total electrostatic energy of point-like particles was calculated that is finite.
We have obtained the correction to Coulomb's law at $r\rightarrow \infty$ that is in the order of ${\cal O}(r^{-6})$.
The dilatation symmetry is broken because of the presence of the dimensional
parameters $\beta$ and $\gamma$.
It was shown that WEC, DEC and SEC are satisfied for the case $\textbf{B}=0$, $\textbf{E}\neq0$. If $\textbf{E}=0$, $\textbf{B}\neq0$ WEC and DEC hold for any values of the magnetic field but SEC is satisfied at $B\leq \sqrt{2}/\sqrt{3\beta}$. This means that magnetized universe is accelerating when the average magnetic field $B\geq \sqrt{2}/\sqrt{3\beta}$.
The electric-magnetic duality between $P$ and $F$ frames is broken in the model under consideration.

We have studied the magnetized black holes in GR and obtained the regular black hole
solution and its asymptotic at $r\rightarrow \infty$.
The magnetic mass of the black hole and the metric function were calculated. We have demonstrated that the Ricci scalar does not possess singularities at $r\rightarrow \infty$ and at $r\rightarrow 0$.
For different value of the parameter $c=2^{9/2}\sqrt{\beta}/(Gq)$ there can be one horizon ($c\simeq 5.625$), two horizons ($c<5.625$) or no horizons ($c>5.625$) corresponding to a particle-like solution.

The thermal stability of regular black holes was investigated and the
Hawking temperature of black holes was evaluated. At $r_+\simeq 1.94\sqrt{q}\beta^{1/4}$ the first-order phase transition takes place and Hawking temperature becomes zero, $T_H=0$. At $r_+<1.94\sqrt{q}\beta^{1/4}$ the black hole is unstable.
The heat capacity diverges at $r_+\simeq 3.92\sqrt{q}\beta^{1/4}$ that indicates on the phase transition of the second-order.
The parameters $\beta$, $m$ and $T_H$ corresponding to first-order and second-order phase transitions were calculated.
In our opinion the model proposed, is of theoretical interest.


\begin{thebibliography}{99}

\bibitem{Heisenberg} W. Heisenberg and H. Euler, Z. Physik, \textbf{98}, 714 (1936) (arXiv:physics/0605038).

\bibitem{Rizzo} A. Cadene, P. Berceau, M. Fouche, R. Battesti and C. Rizzo, Eur. Phys. J. D \textbf{68}, 16 (2014) (arXiv:1302.5389).
\bibitem{Valle} F. Della Valle, et al, Phys. Rev. D \textbf{90} 092003 (2014).

\bibitem{Battesti} R. Battesti and C. Rizzo, Rep. Prog. Phys. \textbf{76}, 016401 (2013) (arXiv:1211.1933).

\bibitem{Born} M. Born and L. Infeld, Proc. Royal Soc. (London) A \textbf{144}, 425 (1934).

\bibitem{Krug} S. I. Kruglov, J. Phys. A \textbf{43}, 375402 (2010) (arXiv:0909.1032).

\bibitem{Jackson} J. D. Jackson, \textit{Classical Electrodynamics, Second Ed.} (John Wiley and Sons, 1975).

\bibitem{Shabad} D. M. Gitman, A. E. Shabad, Eur. Phys. J. C \textbf{74}, 3186 (2014).

\bibitem{Kruglov} S. I. Kruglov, Ann. Phys. \textbf{353}, 299 (2015) (arXiv:1410.0351).

\bibitem{Kruglov2} S. I. Kruglov, Ann. Phys. (Berlin) \textbf{527}, 397 (2015) (arXiv:1410.7633).

\bibitem{Kruglov7} S. I. Kruglov, Commun. Theor. Phys. \textbf{66}, 59 (2016) (arXiv:1511.03303).

\bibitem{Garcia} R. Garc\'{i}a-Salcedo and N. Breton, Int. J. Mod. Phys. A \textbf{15}, 4341 (2000) (arXiv:gr-qc/0004017).

\bibitem{Camara} C. S. Camara, M. R. de Garcia Maia, J. C. Carvalho and J. A. S. Lima, Phys. Rev. D
    \textbf{69}, 123504 (2004) (arXiv:astro-ph/0402311)..
\bibitem{Novello} M. Novello, S. E. Perez Bergliaffa and J. M. Salim, Phys. Rev. D \textbf{69}, 127301
    (2004) (arXiv:astro-ph/0312093).

\bibitem{Novello1} M. Novello, E. Goulart, J. M. Salim and S. E. Perez Bergliaffa, Class. Quant. Grav.
    \textbf{24}, 3021 (2007) (arXiv:gr-qc/0610043).

\bibitem{Vollick} D. N. Vollick, Phys. Rev. D \textbf{78}, 063524 (2008) (arXiv:0807.0448).

\bibitem{Kruglov3} S. I. Kruglov, Phys. Rev. D \textbf{92}, 123523 (2015) (arXiv:1601.06309).

\bibitem{Kruglov0} S. I. Kruglov,  Int. J. Mod. Phys. A \textbf{31}, 1650058 (2016) ( arXiv:1607.03923).

\bibitem{Kruglov4} S. I. Kruglov, Int. J. Mod. Phys. D \textbf{25}, 1640002 (2016) (arXiv:1603.07326).

\bibitem{Quiros} R. Garc\'{i}a-Salcedo, T. Gonzalez and I. Quiros, Phys. Rev. D \textbf{89}, 084047 (2014) (arXiv: 1312.3163).

\bibitem{Oliveira} H. P. de Oliveira, Class. Quant. Grav. \textbf{11}, 1469 (1994).

\bibitem{Soleng} H. H. Soleng, Phys. Rev. D \textbf{52}, 6178 (1995) (arXiv:hep-th/9509033).

\bibitem{Breton} N. Breton, Phys. Rev. D \textbf{67}, 124004 (2003) (arXiv:hep-th/0301254).

\bibitem{Lemos} J. P. S. Lemos and V. T. Zanchin, Phys. Rev. D \textbf{83}, 124005 (2011) (arXiv:1104.4790).

\bibitem{Hendi} S. H. Hendi, Ann. Phys. \textbf{333}, 282 (2013) (arXiv:1405.5359).

\bibitem{Balard} L. Balart and E. C. Vagenas, Phys. Rev. D \textbf{90}, 124045 (2014) (arXiv:1408.0306).

\bibitem{Kruglov5} S. I. Kruglov, Int. J. Geom. Meth. Mod. Phys. \textbf{12}, 1550073 (2015) (arXiv:1504.03941).

\bibitem{Kruglov6} S. I. Kruglov, Ann. Phys. (Berlin) \textbf{528}, 588 (2016) (arXiv:1607.07726).

\bibitem{Kruglov8} S. I. Kruglov, Phys. Rev. D \textbf{94}, 044026 (2016) (arXiv:1608.04275).

\bibitem{Bardeen} J. Bardeen, in: Proceedings of GR5, Tbilisi, USSR, 1968, p. 174.

\bibitem{Ayon} E. Ay\'{o}n-Beato, A. Gar\'{c}ia, Phys. Lett. B \textbf{493}, 149 (2000).

\bibitem{Ayon1} E. Ay\'{o}n-Beato, A. Gar\'{c}ia, Phys. Rev. Lett.  \textbf{80}, 5056 (1998) (arXiv:gr-qc/9911046).

\bibitem{Ayon2} E. Ay\'{o}n-Beato, A. Gar\'{c}ia, Gen. Relat. Grav.  \textbf{31}, 629 (1999).

\bibitem{Ayon3} E. Ay\'{o}n-Beato, A. Gar\'{c}ia, Phys. Lett. B \textbf{464}, 25 (1999).

\bibitem{Bronnikov} K. A. Bronnikov, Phys. Rev. D \textbf{63}, 044005 (2001).

\bibitem{Dymnikova} I. Dymnikova, Class. Quant. Grav., \textbf{21}, 4417 (2004).

\bibitem{Kruglov9} S. I. Kruglov, Phys. Rev. D \textbf{75}, 117301 (2007).

\bibitem{Dittrich} W. Dittrich and H. Gies, Springer Tracts Mod. Phys. \textbf{166}, 1 (2000).

\bibitem{Shabad2}A. E. Shabad, V. V. Usov, Phys. Rev. D \textbf{83}, 105006 (2011) (arXiv:1101.2343).

\bibitem{Hehl} F. W. Hehl and Yu. N. Obukhov, \textit{Foundations of classical electrodynamics: Chage, flux, and metric} (Birkh\"{a}user, Boston, 2003).

\bibitem{Gibbons} G. W. Gibbons and D. Rasheed,  Nucl. Phys. B \textbf{454} (1995) 185 (arXiv:hep-th/9506035).
.
\bibitem{Landau} L. D. Landau and E. M. Lifshits, \textit{The classical theory of fields} (Pergamon Press, 1975).

\bibitem{Born1} M. Born and L. Infeld, Nature \textbf{132} (1933) 970.

\bibitem{Rohrlich} F. Rohrlich, \textit{Classical Charged Particles} (AddisonWesley, Redwood City, CA, 1990).

\bibitem{Spohn} H. Spohn, \textit{Dynamics of Charged Particles and Their Radiation Field} (Cambridge University Press, Cambridge, 2004).

\bibitem{Dirac1} P. A. M. Dirac, Proc. Royal Soc. (London) A \textbf{268} (1962) 57.

\bibitem{Garcia1} H. Salazar, A. Garc\'{i}a, and J. Pleba\'{n}ski, J. Math. Phys., \textbf{28}, 2171 (1987).

\bibitem{Hawking} S. W. Hawking and G. F. R. Ellis, \textit{The Large Scale Structure of Space-Time}, (Cambridge Univ. Press, 1973).

\bibitem{Bardeen1} J. M. Bardeen, Phys. Rev. Lett. \textbf{46}, 382 (1981).

\bibitem{Israel} W. Israel, Phys. Rev. Lett. \textbf{57}, 397 (1986).

\end{thebibliography}
\end{document}